\documentstyle[preprint,prb,aps]{revtex}

\begin{document}
\draft
\title{Understanding light quanta:\\
Construction of the free electromagnetic field}
\author{A. C. de la Torre}
\address{Departamento de F\'{\i}sica,
 Universidad Nacional de Mar del Plata\\
 Funes 3350, 7600 Mar del Plata, Argentina\\
dltorre@mdp.edu.ar}
\maketitle
 \begin{center}
quant-ph/0503023\end{center}
\begin{abstract}
The free electromagnetic field, solution of Maxwell's equations
and carrier of energy, momentum and spin, is construed as an
emergent collective property of an ensemble of photons, and with
this, the consistency of an interpretation that considers that
the photons, and not the electromagnetic fields, are the primary
ontology is established.
\end{abstract}

\section{INTRODUCTION}
This work is the third of a series of papers devoted to the
construction and presentation of a consistent understanding of
the quantum mechanical description of electromagnetic radiation,
based on the assumption that the photons, and not the
electromagnetic fields, constitute the basic ontology; that is,
the photons are the objectively existent building blocks that
require a quantum mechanical treatment and the electromagnetic
fields are an emergent collective property of an ensemble of
photons (as is, for instance, the pressure or temperature of an
ensemble of molecules) and therefore a direct quantization of the
fields is not necessary. This assumption may seem trivial and
unnecessary to many physicists however the opposite view has been
also adopted, namely, that the electromagnetic fields are the
really existent objects and the photons are just a mathematical
tool or ``particle-like excitations''\cite{bal} (quasi particles)
corresponding to the normal modes of oscillations of the fields
(in a similar way that phonons are used to describe lattice
vibrations)

In the first paper of this series\cite{phot1}, the commutation
relations of the operators associated to the free electromagnetic
field were derived from first principles (without reference to
quantum field theory) and their singular character was analysed.
The conclusion reached there is that we should not consider the
electromagnetic fields as a primary ontology that must be treated
quantum mechanically but instead, they can be thought to be a
collective manifestation emerging from an ensemble of fundamental
entities -the photons- with objective existence. One of the
reasons for considering that the photons, and not the
electromagnetic fields, are more appropriate for the quantum
description of the electromagnetic phenomena is that the
quantization of the lagrangian field theory for electromagnetism
presents several difficulties, that can of course be solved, but
are indicative that the electromagnetic fields are perhaps not
the best  language for the quantum mechanical description of
electromagnetic radiation. One of these difficulties is, for
example, that the canonical field variables (the four-vector
potential) an its corresponding conjugate momentum are subject to
several constraints\cite{wei}. Another example is that, in order
to maintain relativistic covariance, every Lorentz transformation
must be accompanied by a gauge transformation\cite{Yndu}. Much
simplicity is gained if we accept the objective existence of the
photons as the elementary entities responsible for the
electromagnetic phenomena. These photons, massless relativistic
particles, were the main concern of the second paper\cite{phot2}.
They are treated by quantum mechanics in the usual way except for
the complications arising from their massless character that
imposes a coupling of the spin states with the momentum states.
For this reason there is a clear preference for treating the
photons in their momentum eigenstates as compared with localized
states. The \emph{elements of physical reality} of the classical
relativistic photon were formalized in the second paper of the
series by the definition of a \emph{photon tensor} $f^{\mu\nu}$,
that is most conveniently described in terms of the three-vectors
${\mathbf e}$ and ${\mathbf b}$. However these quantities
\emph{are not} the space components of a four-vector because they
do not transform as such in a general Lorentz transformation.
They should be considered as a convenient notation for the six
nonvanishing components of the photon tensor $f^{\mu\nu}$,
according to the assignment given by
\begin{equation}\label{fmunu}
   f^{\mu\nu}= \left(%
\begin{array}{cccc}
  0 & e_{1} & e_{2} & e_{3} \\
  -e_{1} & 0 & b_{3} & -b_{2} \\
  -e_{2} & -b_{3} & 0 & b_{1} \\
  -e_{3} & b_{2} & -b_{1} & 0 \\
\end{array}%
\right)\ .
\end{equation}
The energy $E$, momentum ${\mathbf P}$ and spin ${\mathbf S}$ of
a photon are related by
\begin{equation}\label{ep}
 E=c|{\mathbf P}|
\ ,\
 {\mathbf S}\times {\mathbf P} =0\ ,
\end{equation}
and as a consequence of its massless character (not of quantum
mechanics), the energy must be related to some intrinsic
frequency $\omega$. \emph{In any reference frame, we can
visualize a positive or negative helicity photon of energy $E$
and spin $\hbar$, propagating with speed $c$ in a direction given
by a unit vector ${\mathbf k}$, as a unit vector ${\mathbf
\hat{e}}$ rotating clockwise or counterclockwise in a plane
orthogonal to ${\mathbf k}$ with frequency $\omega=E/\hbar$. In
the same plane we have another unit vector ${\mathbf
\hat{b}}={\mathbf k}\times{\mathbf \hat{e}}$ and with the vectors
${\mathbf e}=\omega{\mathbf \hat{e}}$ and ${\mathbf
b}=\omega{\mathbf \hat{b}}$ we can build the photon tensor
$f^{\mu\nu}$ whose Lorentz transformations provide the
description of the photon in other reference frames.}

The rotating vectors ${\mathbf e}_{s}(t)$, corresponding to a
photon of helicity $s=\pm 1$, used to define the photon tensor
can be given more conveniently in terms of the circular
polarization \emph{complex} vectors
$\mbox{\boldmath$\epsilon$}_{s}$ defined by
\begin{eqnarray}
\mbox{\boldmath$\epsilon$}_{+}&=&
 \frac{1}{\sqrt{2}}({\mathbf \hat{e}}  + i {\mathbf \hat{b}}) \\
 \mbox{\boldmath$\epsilon$}_{-}&=&
   \frac{1}{\sqrt{2}}(i {\mathbf \hat{e}}  + {\mathbf \hat{b}})
\end{eqnarray}
resulting in
\begin{eqnarray}
 {\mathbf e}_{+}(t)&=&
 \omega ({\mathbf \hat{e}} \cos \omega t + {\mathbf \hat{b}}\sin \omega t)
 \ =\ \left(\frac{\omega}{\sqrt{2}}\ \mbox{\boldmath$\epsilon$}_{+}{\mathrm e}^{-i\omega t} +
  {\mathrm c.c.}\right)\\
  {\mathbf e}_{-}(t) &=&
  \omega ({\mathbf \hat{b}} \cos \omega t + {\mathbf \hat{e}}\sin \omega t)
  \ =\ \left(\frac{\omega}{\sqrt{2}}\ \mbox{\boldmath$\epsilon$}_{-}{\mathrm e}^{-i\omega t} +
  {\mathrm c.c.}\right)\ ,
\end{eqnarray}
where c.c. stands for complex conjugation of the previous term.
(Another initial position of the vector can be achieved simply by
multiplying the circular polarization complex vectors by a phase,
that is, ${\mathrm e}^{-i\theta}\mbox{\boldmath$\epsilon$}_{s}$).
We have then
\begin{equation}\label{eveceigval}
   {\mathbf e}_{s}(t)=
 \left(\frac{\omega}{\sqrt{2}}\ \mbox{\boldmath$\epsilon$}_{s}
 {\mathrm e}^{-i\omega t} +
  {\mathrm c.c.}\right)\ .
\end{equation}
The other vector, ${\mathbf b}_{s}(t)$, needed in order to build
the photon tensor is simply obtained as ${\mathbf
b}_{s}(t)={\mathbf k}\times{\mathbf e}_{s}(t)$. The circular
polarization complex vectors depend, of course, on the direction
of propagation of the photon, ${\mathbf k}$, and could be denoted
by $\mbox{\boldmath$\epsilon$}_{s}({\mathbf k})$; however, for
simplicity, we will not show this dependence explicitly in the
notation. The usual orthogonality relations are
\begin{eqnarray}
  \mbox{\boldmath$\epsilon$}^{\ast}_{s}\cdot{\mathbf k}&=& 0\ , \\
  \mbox{\boldmath$\epsilon$}^{\ast}_{s}\cdot\mbox{\boldmath$\epsilon$}_{s'} &=&
  \delta_{s,s'}\ , \\
  \mbox{\boldmath$\epsilon$}^{\ast}_{s}\times\mbox{\boldmath$\epsilon$}_{s'}&=&
  si{\mathbf k}\delta_{s,s'} \ ,\\
 {\mathbf k} \times\mbox{\boldmath$\epsilon$}_{s}&=&
  s \mbox{\boldmath$\epsilon$}^{\ast}_{-s}\ ,
\end{eqnarray}
and for further reference we present some useful algebraic
relations
\begin{eqnarray}
  \mbox{\boldmath$\epsilon$}_{-}&=& i\ \mbox{\boldmath$\epsilon$}^{\ast}_{+}\ , \\
  \mbox{\boldmath$\epsilon$}_{s}\cdot\mbox{\boldmath$\epsilon$}_{s'} &=&
  i(1-\delta_{s,s'}) = i\delta_{s,-s'}\ ,  \\
  \mbox{\boldmath$\epsilon$}_{s}\times\mbox{\boldmath$\epsilon$}_{s'}&=&
  s {\mathbf k}(1- \delta_{s,s'} )=  s {\mathbf k}\delta_{s,-s'} \
  .
\end{eqnarray}
Another relation that will be later needed is
\begin{equation}\label{helicity summ}
  \sum_{s=\pm} (\mbox{\boldmath$\epsilon$}^{\ast}_{s})_{i}\
  (\mbox{\boldmath$\epsilon$}_{s})_{j} = \delta_{i,j}-({\mathbf k})_{i}({\mathbf
  k})_{j}\ ,
\end{equation}
where $(\mbox{\boldmath$\epsilon$}_{s})_{j}$ and $({\mathbf
k})_{j}$ are the cartesian components of the corresponding
vectors in an arbitrary set of orthogonal unit vectors $({\mathbf
\hat{x}}_{1},{\mathbf \hat{x}}_{2},{\mathbf \hat{x}}_{3})$. In
order to prove this, notice first that $ \sum_{s=\pm}
(\mbox{\boldmath$\epsilon$}^{\ast}_{s})_{i}\
  (\mbox{\boldmath$\epsilon$}_{s})_{j} = ({\mathbf \hat{e}})_{i}({\mathbf
  \hat{e}})_{j} +({\mathbf \hat{b}})_{i}({\mathbf \hat{b}})_{j}$.
  Now, $({\mathbf
\hat{e}})_{i},({\mathbf \hat{b}})_{i},({\mathbf k})_{i}$ are the
cartesian components of ${\mathbf \hat{x}}_{i}$ in the orthogonal
set $({\mathbf \hat{e}},{\mathbf \hat{b}},{\mathbf k})$ and
therefore, from the scalar product ${\mathbf
\hat{x}}_{i}\cdot{\mathbf \hat{x}}_{j}=\delta_{i,j}=({\mathbf
\hat{e}})_{i}({\mathbf \hat{e}})_{j}+({\mathbf
\hat{b}})_{i}({\mathbf \hat{b}})_{j}+({\mathbf k})_{i}({\mathbf
k})_{j}$ follows the proof.

The quantum mechanical description of a photon in a Hilbert space
${\mathcal H}={\mathcal H}^{S}\otimes{\mathcal H}^{K}$
corresponding to the spin and kinematic part, is most
conveniently done in terms of eigenstates of fixed helicity
$s=\pm 1$ and momentum ${\mathbf p}$ (in the direction of the
unit vector ${\mathbf k}$), denoted by $\varphi_{s,{\mathbf
p}}=\chi_s^{{\mathbf k}}\otimes\phi_{{\mathbf p}}$. For an
explicit representation of these states we can choose for the
three dimensional spin-Hilbert space ${\mathcal H}^{S}$, the one
where the spin operators take the matrix form
\begin{equation}\label{spinmat}
    S_{x}=\hbar\left(%
\begin{array}{rrr}
  0 & 0 & 0 \\
  0 & 0 & -i \\
  0 & i& 0 \\
\end{array}%
\right)\ ,\
  S_{y}=\hbar\left(%
\begin{array}{rrr}
  0 & 0 & i \\
  0 & 0 & 0 \\
  -i & 0& 0 \\
\end{array}%
\right)\ ,\
  S_{z}=\hbar\left(%
\begin{array}{rrr}
  0 & -i & 0 \\
  i & 0 & 0 \\
  0 & 0 & 0 \\
\end{array}%
\right)\ ;
\end{equation}
that is, with the matrix elements given by
\begin{equation}\label{spinmat1}
  ( S_{j} )_{kl}=-i\hbar\varepsilon_{jkl}\ .
\end{equation}
In this space, the helicity states are
\begin{equation}\label{eigev}
   \chi_{\pm}^{{\mathbf k}}=\frac{1}{2\sqrt{1-k_{x}k_{y}-k_{y}k_{z}-k_{z}k_{x}}}\left(%
\begin{array}{c}
  1- k_{x}(k_{x}+k_{y}+k_{z})\pm i(k_{y}-k_{z})\\
1- k_{y}(k_{x}+k_{y}+k_{z})\pm i(k_{z}-k_{x})\\
1- k_{z}(k_{x}+k_{y}+k_{z})\pm i(k_{x}-k_{y}) \\
\end{array}%
\right)\ .
 \end{equation}
For the kinematic description we can choose the space of
 square integrable functions (more precisely, its \emph{rigged}
 extension) in the position representation, where the momentum eigenstates are given by
 \begin{equation}\label{posrep}
\phi_{{\mathbf p}}({\mathbf
r})=\frac{1}{(\sqrt{2\pi}\hbar)^{3}}\exp(\frac{i}{\hbar}{\mathbf
p}\cdot{\mathbf r}) \ .
\end{equation}

A free photon with fixed helicity and momentum is then described
by the state $\varphi_{s,{\mathbf p}}=\chi_s^{{\mathbf
k}}\otimes\phi_{{\mathbf p}}$, and we can use the representations
given in Eqs.(\ref{eigev}) and( \ref{posrep}) for any explicit
calculation. These states build a basis suitable for the
construction of any arbitrary state for one single photon.
However the most interesting physical systems involve a large, or
undetermined, number of photons whose state is presented in next
section.

After having summarized the main results of the previous papers
we will see in this paper how the electromagnetic fields are
built and emerge as an observable property of an ensemble of
photons.
\section{FOCK SPACE STATES FOR MANY PHOTONS}
The most effective way of dealing with a quantum system
consisting in many photons, or with an indefinite number of
photons, is to define the Fock space of states for the system.
The Fock space is built by the orthogonal sum of the vacuum space
plus the Hilbert space for one photon, plus the Hilbert space for
two photons and so on. In this space we define the operators
$a^{\dag}_{s}({\mathbf p})$ and $a_{s}({\mathbf p})$
corresponding to the creation or annihilation of a photon with
momentum ${\mathbf p}$ and helicity $s=\pm 1$. The effect of
these operators on a state for a system with $n$ photons having
helicity and momenta $s_{1}{\mathbf p}_{1},s_{2}{\mathbf
p}_{2},\cdots, s_{n}{\mathbf p}_{n}$ are given by

\begin{equation}\label{a+f}
a^{\dag}_{s}({\mathbf p})\ \varphi_{s_{1}{\mathbf p}_{1}, \cdots,
s_{n}{\mathbf p}_{n}} = \sqrt{n+1}\ \varphi_{s{\mathbf
p},s_{1}{\mathbf p}_{1}, \cdots, s_{n}{\mathbf p}_{n}} \ ,
\end{equation}
\begin{equation}\label{a-f}
a_{s}({\mathbf p})\ \varphi_{s_{1}{\mathbf p}_{1}, \cdots,
s_{n}{\mathbf p}_{n}} = \frac{1}{\sqrt{n}}\
\sum_{i=1}^{n}\delta_{s,si}\delta({\mathbf p}-{\mathbf p}_{i})\
\varphi_{s_{1}{\mathbf p}_{1},\cdots, \neg(s_{i}{\mathbf p}_{i}),
\cdots, s_{n}{\mathbf p}_{n}} \ ,
\end{equation}
where the symbol $\neg(s_{i}{\mathbf p}_{i})$ indicates that the
corresponding indices are eliminated if they are present. The
vacuum state $\varphi_{0}$ with zero photons is such that
\begin{equation}\label{vacuum}
a_{s}({\mathbf p})\ \varphi_{0} = 0\ ,
\end{equation}
and an $n$ photon state is built applying the creation operator
to the vacuum state,
\begin{equation}\label{n phot st}
\varphi_{s_{1}{\mathbf p}_{1},s_{2}{\mathbf p}_{2}, \cdots,
s_{n}{\mathbf p}_{n}}=
\frac{1}{\sqrt{n!}}a^{\dag}_{s_{1}}({\mathbf
p_{1}})a^{\dag}_{s_{2}}({\mathbf p_{2}})\cdots
a^{\dag}_{s_{n}}({\mathbf p_{n}}) \varphi_{0} \ .
\end{equation}
The symmetry requirements for identical boson states impose the
commutation relations for the creation and annihilation
operators\begin{equation}\label{com rel}
   [a_{s}({\mathbf
p}),a^{\dag}_{s'}({\mathbf p'})]=
\delta_{s,s'}\delta^{3}({\mathbf p}-{\mathbf p'})\ ,\
[a^{\dag}_{s}({\mathbf p}),a^{\dag}_{s'}({\mathbf p'})]=
[a_{s}({\mathbf p}),a_{s'}({\mathbf p'})]= 0 \ .
\end{equation}
Finally, the operator corresponding to the number of photons with
helicity $s$ and with momentum within $d^{3}{\mathbf p}$ centered
in ${\mathbf p}$ is given by
\begin{equation}\label{nr of fot}
    N_{s}({\mathbf
p})=a^{\dag}_{s}({\mathbf p})a_{s}({\mathbf p})\ ,
\end{equation}
and the operator for the total number of photons in the system is
\begin{equation}\label{tot nr phot}
   N=\sum_{s}\int\!\!\! d^{3}{\mathbf p}\ N_{s}({\mathbf
p}) \ .
\end{equation}

The creation and annihilation operators are not only useful for
the representation of the state of many photons but they can also
be used to represent \emph{any} observable of a multi-photon
system. This is so because any operator can be given in terms of
the spectral decomposition involving projectors in their
eigenstates. Now if we represent these eigenstates in terms of
the Fock basis generated by the application of creations
operators to the vacuum state, we finally obtain the operator
expressed with annihilation and creation operators. The fact that
every observable can be given in terms of creation and
annihilation operators will be relevant in next section where we
will try to discover relevant observables by considering the
simplest construction of hermitian operators with the
nonhermitian operators $a^{\dag}_{s}({\mathbf p})$ and
$a_{s}({\mathbf p})$.
\section{MANY PHOTON OBSERVABLES AND THE CONSTRUCTION OF THE ELECTROMAGNETIC FIELD}
If we accept that the photons have objective existence and that
each of them carries  momentum $\mathbf{p}$,  energy
$E=c|{\mathbf p}|=\hbar\omega$ and spin $\pm \hbar$ in the
direction of propagation $\mathbf{k}$, we can build the total
momentum, energy and spin of a system of many non interacting
photons, simply as the sum of the corresponding contribution of
each photon. The total energy, momentum and spin are then
\begin{equation}\label{tot.ener}
   H=\sum_{s}\int\!\!\! d^{3}{\mathbf p}\ \hbar\omega\ N_{s}({\mathbf
p})\ ,
\end{equation}
\begin{equation}\label{tot.mom}
   {\mathbf P}=\sum_{s}\int\!\!\! d^{3}{\mathbf p}\ {\mathbf p}\ N_{s}({\mathbf
p})\ ,
\end{equation}
\begin{equation}\label{tot.spin}
   {\mathbf S}=\int\!\!\! d^{3}{\mathbf p}\ \hbar{\mathbf k}\ \left(N_{+}({\mathbf
p})-N_{-}({\mathbf p})\right)\ .
\end{equation}
The assumption that the photons are non interacting is a very
good approximation because photons couple only to charged
particles and the leading contribution photon-photon interaction,
corresponding to a ``box graph'' Feynman diagram, is of
\emph{fourth} order in perturbation theory and can therefore be
ignored. All the observables above are given in terms of the
number operator $N_{s}({\mathbf p})=a^{\dag}_{s}({\mathbf
p})a_{s}({\mathbf p})$; however we can expect that besides this
number operator there is another relevant hermitian operator
related to the creation and annihilation operator. This
expectation comes from the fact that a non hermitian operator
like $a^{\dag}_{s}({\mathbf p})$ or $a_{s}({\mathbf p})$ is
related to \emph{two} independent hermitian operators (this is
similar to the case of complex numbers that contain \emph{two}
independent real numbers). Now, the number operator is just one
of them (actually it is the operator modulus squared of
$a_{s}({\mathbf p})$) and we can expect the existence of another
relevant operator, or observable of a multi-photon system,
corresponding to the hermitian or antihermitian part of
$a_{s}({\mathbf p})$. Therefore we expect an observable of the
form
\begin{equation}\label{observ form}
\sum_{s}\int\!\!\! d^{3}{\mathbf p}\ \left(f(s,{\mathbf
p},E,{\mathbf r},t)\ a_{s}({\mathbf p}) \pm {\mathrm
h.c.}\right)\ ,
\end{equation}
where h.c stands for ``hermitian conjugate'' of the previous
term. Notice that the spin, momentum and energy of each
individual photon are integrated and therefore this observable is
related to the ensemble of photons as a collective property. Now
we can make some considerations in order to make an educated
guess of the form that the function $f(s,{\mathbf p},E,{\mathbf
r},t)$ can take. First, this function must involve the elements
of physical reality of the contributing photons that are
formalized by the two orthogonal rotating vectors ${\mathbf
e}_{s}(t)$, given in Eq.(\ref{eveceigval}), and ${\mathbf
b}_{s}(t)={\mathbf k}\times{\mathbf e}_{s}(t)$; therefore we have
two choices: $\mbox{\boldmath$\epsilon$}_{s} {\mathrm
e}^{-i\omega t}$ and $({\mathbf
k}\times\mbox{\boldmath$\epsilon$}_{s}) {\mathrm e}^{-i\omega
t}$. Next we can expect that the space dependence of the function
will be the same as the space dependence of the photon state in
the position representation as given in Eq.(\ref{posrep}). For
the energy dependence, we don't have any argument suggesting a
particular form. Consistent with all this, we propose the two
hermitian operators
\begin{equation}\label{Efield}
{\mathbf E}({\mathbf r},t) =\frac{1}{2\pi\hbar}\
\sum_{s}\int\!\!\! d^{3}{\mathbf p}\ \sqrt{\omega}\left(i\
a_{s}({\mathbf p})\ \mbox{\boldmath$\epsilon$}_{s}\ {\mathrm
e}^{\frac{i}{\hbar}({\mathbf p}\cdot{\mathbf r}-Et)} + {\mathrm
h.c.}\right)\ ,
\end{equation}
\begin{equation}\label{Bfield}
{\mathbf B}({\mathbf r},t) =\frac{1}{2\pi\hbar}\
\sum_{s}\int\!\!\! d^{3}{\mathbf p}\ \sqrt{\omega}\left(i\
a_{s}({\mathbf p})\ ({\mathbf
k}\times\mbox{\boldmath$\epsilon$}_{s})\ {\mathrm
e}^{\frac{i}{\hbar}({\mathbf p}\cdot{\mathbf r}-Et)} + {\mathrm
h.c.}\right)\ .
\end{equation}
The symbols used to denote these operators suggest that they
correspond to the electromagnetic fields but, of course, we
haven't yet given any argument supporting this. That these
operators are indeed the electromagnetic fields, will be
established when we derive relations among them and with the
total energy, momentum and spin. Before doing this, it is
convenient to define another operator such that its time and
space derivatives result in the two operators above. This is,
\begin{equation}\label{Afield}
{\mathbf A}({\mathbf r},t) =\frac{c}{2\pi\hbar}\
\sum_{s}\int\!\!\! d^{3}{\mathbf p}\
\frac{1}{\sqrt{\omega}}\left(\ a_{s}({\mathbf p})\
\mbox{\boldmath$\epsilon$}_{s}\ {\mathrm
e}^{\frac{i}{\hbar}({\mathbf p}\cdot{\mathbf r}-Et)} + {\mathrm
h.c.}\right)\ ,
\end{equation}
and we have
\begin{equation}\label{AtoEB}
{\mathbf E}({\mathbf r},t)=-\frac{1}{c}\partial_{t}{\mathbf
A}({\mathbf r},t)\ ,\ {\mathbf B}({\mathbf
r},t)=\nabla\times{\mathbf A}({\mathbf r},t)\ .
\end{equation}

The operators in Eqs.(\ref{Efield}, \ref{Bfield}, \ref{Afield})
have linear dependence on the creation and annihilation operators
whereas the total energy, momentum and spin given in
Eqs.(\ref{tot.ener}, \ref{tot.mom}, \ref{tot.spin}) have a
quadratic dependence on them. Therefore we can expect that the
total energy, momentum and spin will be related to \emph{products} of
the operators above. In fact, we can show that
\begin{equation}\label{HtoEB}
 H=\frac{1}{8\pi}\int\!\!\! d^{3}{\mathbf r}\ \left({\mathbf E}^{2}+{\mathbf
B}^{2}\right)\ ,
\end{equation}
\begin{equation}\label{PtoEB}
{\mathbf P}=\frac{1}{8\pi c}\int\!\!\! d^{3}{\mathbf r}\
\left({\mathbf E}\times{\mathbf B}-{\mathbf B}\times{\mathbf
E}\right)\ ,
\end{equation}
\begin{equation}\label{StoEA}
{\mathbf S}=\frac{1}{8\pi c}\int\!\!\! d^{3}{\mathbf r}\
\left({\mathbf E}\times{\mathbf A}-{\mathbf A}\times{\mathbf
E}\right)\ .
\end{equation}
These are the usual expressions for the energy, momentum and spin
of the electromagnetic fields found in any electrodynamic text,
except that here we allow for the possible non-commutation of the
fields. The time dependence of the fields in Eqs.(\ref{Efield},
\ref{Bfield}, \ref{Afield}) is cancelled in the combinations of
the integrands in Eqs.(\ref{HtoEB}, \ref{PtoEB}, \ref{StoEA}) and
the resulting operators are conserved. The proof that the
integrals in Eqs.(\ref{HtoEB}, \ref{PtoEB}, \ref{StoEA}) lead to
the operators in Eqs.(\ref{tot.ener}, \ref{tot.mom},
\ref{tot.spin}) is given in the appendix.

As a final confirmation that the fields in
Eqs.(\ref{Efield},\ref{Bfield}) are indeed the free
electromagnetic fields, we can see that they satisfy Maxwell's
equations.
\begin{eqnarray}
 -\nabla\times{\mathbf E}&=& \frac{1}{c}\partial_{t}{\mathbf B}\ , \\
\nabla\times{\mathbf B}&=& \frac{1}{c}\partial_{t}{\mathbf E}\ ,\\
\nabla\cdot{\mathbf E}&=& 0\ ,\\
\nabla\cdot{\mathbf B}&=& 0\ .
\end{eqnarray}
We have indeed,
\begin{eqnarray}
 -\nabla\times{\mathbf E}&=& -\frac{1}{2\pi\hbar}\
\sum_{s}\int\!\!\! d^{3}{\mathbf p}\ \sqrt{\omega}\left(i\
a_{s}({\mathbf p})\ \frac{i}{\hbar}({\mathbf
p}\times\mbox{\boldmath$\epsilon$}_{s})\ {\mathrm
e}^{\frac{i}{\hbar}({\mathbf p}\cdot{\mathbf r}-Et)} +
{\mathrm h.c.}\right) \nonumber \\
&=& \frac{1}{2\pi\hbar}\ \sum_{s}\int\!\!\! d^{3}{\mathbf p}\
\sqrt{\omega}\left(\ a_{s}({\mathbf p})\
\frac{E}{c\hbar}({\mathbf
k}\times\mbox{\boldmath$\epsilon$}_{s})\ {\mathrm
e}^{\frac{i}{\hbar}({\mathbf p}\cdot{\mathbf r}-Et)} +
{\mathrm h.c.}\right) \nonumber \\
&=& \frac{1}{c2\pi\hbar}\ \sum_{s}\int\!\!\! d^{3}{\mathbf p}\
\sqrt{\omega}\left(\ a_{s}({\mathbf p})\ ({\mathbf
k}\times\mbox{\boldmath$\epsilon$}_{s})\ i\partial_{t}{\mathrm
e}^{\frac{i}{\hbar}({\mathbf p}\cdot{\mathbf r}-Et)} +
{\mathrm h.c.}\right) \nonumber \\
&=&\frac{1}{c}\partial_{t}{\mathbf B} \ ,
\end{eqnarray}
and the other equations are proved in similar way.

With the fields given in terms of the creation and annihilation
operators, we can now prove the commutation relations that were
derived in reference (2) from general principles. Let us
calculate first the commutation relations among two cartesian
components of the electric field.
\begin{eqnarray}
 [E_{i}({\mathbf r}_{1},t_{1})&,&E_{j}({\mathbf r}_{2},t_{2})]=
 \frac{1}{(2\pi\hbar)^{2}}\ \sum_{s} \sum_{s'}\int\!\!\!
d^{3}{\mathbf p} \int\!\!\! d^{3}{\mathbf p'}\
\sqrt{\omega\omega'}\nonumber \\
& &\left[ \left( i\ a_{s}({\mathbf p})\
(\mbox{\boldmath$\epsilon$}_{s})_{i}\ {\mathrm
e}^{\frac{i}{\hbar}({\mathbf p}\cdot{\mathbf r_{1}}-Et_{1})} - i\
a^{\dag}_{s}({\mathbf p})\ (\mbox{\boldmath$\epsilon$}^{\ast}_{s})_{i}\
{\mathrm e}^{\frac{-i}{\hbar}({\mathbf p}\cdot{\mathbf
r_{1}}-Et_{1})} \right)\right.
, \nonumber \\
& & \left. \left( i\ a_{s'}({\mathbf p'})\
(\mbox{\boldmath$\epsilon$}_{s'})_{j}\ {\mathrm
e}^{\frac{i}{\hbar}({\mathbf p'}\cdot{\mathbf r_{2}}-E't_{2})} -
i\ a^{\dag}_{s'}({\mathbf p'})\
(\mbox{\boldmath$\epsilon$}^{\ast}_{s'})_{j}\ {\mathrm
e}^{\frac{-i}{\hbar}({\mathbf p'}\cdot{\mathbf
r_{2}}-E't_{2})}\right) \right] \nonumber \\
&=& \frac{1}{(2\pi\hbar)^{2}}\ \sum_{s} \sum_{s'}\int\!\!\!
d^{3}{\mathbf p} \int\!\!\! d^{3}{\mathbf p'}\
\sqrt{\omega\omega'} \left( \left[a_{s}({\mathbf p}),
a^{\dag}_{s'}({\mathbf p'})
\right](\mbox{\boldmath$\epsilon$}_{s})_{i}
(\mbox{\boldmath$\epsilon$}^{\ast}_{s'})_{j} {\mathrm
e}^{\frac{i}{\hbar}({\mathbf p}\cdot{\mathbf
r_{1}}-Et_{1})-\frac{i}{\hbar}({\mathbf p'}\cdot{\mathbf
r_{2}}-E't_{2})}\right. \nonumber \\
&+&\left. \left[a^{\dag}_{s}({\mathbf p}), a_{s'}({\mathbf p'})
\right](\mbox{\boldmath$\epsilon$}^{\ast}_{s})_{i}
(\mbox{\boldmath$\epsilon$}_{s'})_{j} {\mathrm
e}^{-\frac{i}{\hbar}({\mathbf p}\cdot{\mathbf
r_{1}}-Et_{1})+\frac{i}{\hbar}({\mathbf p'}\cdot{\mathbf
r_{2}}-E't_{2})}\right) \nonumber \\
&=&\frac{1}{(2\pi\hbar)^{2}}\ \sum_{s} \int\!\!\! d^{3}{\mathbf
p}\ \omega \left((\mbox{\boldmath$\epsilon$}_{s})_{i}
(\mbox{\boldmath$\epsilon$}^{\ast}_{s})_{j} {\mathrm
e}^{\frac{i}{\hbar}\left({\mathbf p}\cdot({\mathbf
r_{1}}-{\mathbf r_{2}})-E(t_{1}-t_{2})\right)}-{\mathrm
c.c.}\right)\nonumber \\
&=&\frac{1}{(2\pi\hbar)^{2}}\ \int\!\!\! d^{3}{\mathbf p}\ \omega
(\delta_{i,j}-({\mathbf k})_{i}({\mathbf
  k})_{j}) \left({\mathrm
e}^{\frac{i}{\hbar}\left({\mathbf p}\cdot({\mathbf
r_{1}}-{\mathbf r_{2}})-E(t_{1}-t_{2})\right)}-{\mathrm
c.c.}\right)\nonumber \\
&=&\frac{-2i}{(2\pi\hbar)^{2}}\ \int\!\!\! d^{3}{\mathbf p}\
\omega (\delta_{i,j}-({\mathbf k})_{i}({\mathbf
  k})_{j})\ {\mathrm
e}^{\frac{i}{\hbar}{\mathbf p}\cdot({\mathbf r_{1}}-{\mathbf
r_{2}})}\ \sin(\omega(t_{1}-t_{2}))\ .
\end{eqnarray}
The last result was obtained doing a variable change ${\mathbf
p}\rightarrow -{\mathbf p}$ in the integration of the ``c.c''
term. This result can be written with help of the singular
function
\begin{eqnarray*}
  D(\mbox{\boldmath$\rho$},\tau) &=& \frac{-1}{(2\pi\hbar)^{3}}\ \int\!\!\! d^{3}{\mathbf p}\
{\mathrm e}^{\frac{i}{\hbar}{\mathbf p}\cdot \mbox{\boldmath$\rho$} }\ \frac{\sin(\omega\tau)}{\omega}\nonumber \\
&=& \frac{-1}{8\pi^{2}c\
\rho}\left[\delta(\rho-c\tau)-\delta(\rho+c\tau)\right]\ ,
\end{eqnarray*}
where $\rho=|\mbox{\boldmath$\rho$}|$. In order to prove the
second expression we can write  the integration over
$d^{3}{\mathbf p}$ in polar coordinates with the ``third'' axis
along the vector $\mbox{\boldmath$\rho$}$. The angle integrations
can be easily performed and the $p=|{\mathbf p}|$ integration
results in the Dirac distributions. In this second representation
we see that the  functions $D(\mbox{\boldmath$\rho$},\tau)$ has
support on the light cone, a fact of physical relevance. With
this, the commutator for the cartesian components of the electric
field is
\begin{equation}\label{ComEE}
 [E_{i}({\mathbf r}_{1},t_{1})\ ,\ E_{j}({\mathbf r}_{2},t_{2})]=
 -i4\pi\hbar c^{2}\left(\delta_{i,j}\frac{1}{c^{2}}\partial_{t1}\partial_{t2}
 +\partial_{r1,i}\partial_{r2,j} \right)\
 D({\mathbf r_{1}}-{\mathbf
r_{2}},t_{1}-t_{2})\ .
\end{equation}
For the calculation of the commutator between the cartesian
components of the magnetic field we use the corresponding
polarization vectors but we obtain the same result,
\begin{equation}\label{ComBB}
 [B_{i}({\mathbf r}_{1},t_{1})\ ,\ B_{j}({\mathbf r}_{2},t_{2})]=
 -i4\pi\hbar c^{2}\left(\delta_{i,j}\frac{1}{c^{2}}\partial_{t1}\partial_{t2}
 +\partial_{r1,i}\partial_{r2,j} \right)\
 D({\mathbf r_{1}}-{\mathbf
r_{2}},t_{1}-t_{2})\ ,
\end{equation}
and the remaining commutator is calculated in similar fashion,
\begin{equation}\label{ComEB}
 [E_{i}({\mathbf r}_{1},t_{1})\ ,\ B_{j}({\mathbf r}_{2},t_{2})]=
 i4\pi\hbar c\ \varepsilon_{ijk}\ \partial_{t1}\partial_{r1,k}\
 \ D({\mathbf r_{1}}-{\mathbf r_{2}},t_{1}-t_{2})\ .
\end{equation}
The singular character of these commutators was discussed in the
first paper\cite{phot1} with the conclusion that the treatment of
the electromagnetic fields as quantum mechanical observables is
perhaps not very meaningful. Regardless of its physical
relevance, we can calculate commutation relations among any
observables build with the creation and annihilation operators
using the basic commutators given in Eq.(\ref{com rel}). For
instance we can easily calculate the commutators of the fields
with the total number of photons
\begin{eqnarray}
  \left[{\mathbf E}({\mathbf r},t)\ ,\ N\right]&=& \frac{1}{2\pi\hbar}\
\sum_{s}\int\!\!\! d^{3}{\mathbf p}\ \sqrt{\omega}\left(i\
a_{s}({\mathbf p})\ \mbox{\boldmath$\epsilon$}_{s}\ {\mathrm
e}^{\frac{i}{\hbar}({\mathbf p}\cdot{\mathbf r}-Et)} - {\mathrm
h.c.}\right)\ ,\\
  \left[ {\mathbf B}({\mathbf r},t)\ ,\ N\right]&=& \frac{1}{2\pi\hbar}\
\sum_{s}\int\!\!\! d^{3}{\mathbf p}\ \sqrt{\omega}\left(i\
a_{s}({\mathbf p})\ ({\mathbf
k}\times\mbox{\boldmath$\epsilon$}_{s})\ {\mathrm
e}^{\frac{i}{\hbar}({\mathbf p}\cdot{\mathbf r}-Et)} - {\mathrm
h.c.}\right)\ , \\
 \left[ {\mathbf A}({\mathbf r},t)\ ,\ N\right]&=& \frac{c}{2\pi\hbar}\
\sum_{s}\int\!\!\! d^{3}{\mathbf p}\
\frac{1}{\sqrt{\omega}}\left(\ a_{s}({\mathbf p})\
\mbox{\boldmath$\epsilon$}_{s}\ {\mathrm
e}^{\frac{i}{\hbar}({\mathbf p}\cdot{\mathbf r}-Et)} - {\mathrm
h.c.}\right)\ .
\end{eqnarray}
Notice, as a curiosity, that the fields are related to the
hermitian part of the creation operators and the commutators
above are related to the anti-hermitian part.
\section{ELECTROMAGNETIC FIELDS OF AN ENSEMBLE OF PHOTONS}
In this section we will see some simple examples of the
electromagnetic fields associated to some multi-photon systems.
The quantity that we must calculate is the expectation value of
the fields in the quantum state describing the system of photons.
The first simplest case is the vacuum, with zero photons,
described by the state $\varphi_{0}$. It follows from
Eq.(\ref{vacuum}) that any operator with the form given in
Eq.(\ref{observ form}) will have zero vacuum expectation value;
therefore
\begin{equation}\label{vac expec val}
 \langle\varphi_{0}\ , {\mathbf E}({\mathbf r},t)\
 \varphi_{0}\rangle = \langle\varphi_{0}\ , {\mathbf B}({\mathbf r},t)\
 \varphi_{0}\rangle = 0\ .
\end{equation}
This is of course expected; however what may be surprising is
that the square of the fields have nonvanishing vacuum
expectation values. In the case of the electric field, for
example, the field in Eq.(\ref{Efield}) has two terms and the
square of it will have four terms, but three of them have
vanishing vacuum expectation value due to Eq.(\ref{vacuum}). The
remaining term is
\begin{eqnarray}
\langle\varphi_{0}\ , {\mathbf E}^{2}({\mathbf r},t)\
 \varphi_{0}\rangle &=&
 \frac{1}{(2\pi\hbar)^{2}}\sum_{s} \sum_{s'}\int\!\!\!
d^{3}{\mathbf p} \int\!\!\! d^{3}{\mathbf p'}\nonumber\\ & &
\sqrt{\omega\omega'}\ \mbox{\boldmath$\epsilon$}_{s}\cdot
\mbox{\boldmath$\epsilon$}^{\ast}_{s'}\ {\mathrm
e}^{\frac{i}{\hbar}({\mathbf p}\cdot{\mathbf r}-Et)}\ {\mathrm
e}^{\frac{-i}{\hbar}({\mathbf p'}\cdot{\mathbf r}-E't)}\
\langle\varphi_{0}\ ,a_{s}({\mathbf p})a^{\dag}_{s'}({\mathbf
p'}) \
 \varphi_{0}\rangle
 \ .
\end{eqnarray}
From the commutation relation of the creation operators we have
$a_{s}({\mathbf p})a^{\dag}_{s'}({\mathbf p'}) =
a^{\dag}_{s'}({\mathbf p'})a_{s}({\mathbf p})+ \delta_{ss'}
\delta({\mathbf p}-{\mathbf p'})$ and the first term has
vanishing vacuum expectation value. Considering also the
orthogonality relation of the polarization vectors, we finally
get the nonzero, even diverging, vacuum expectation value
\begin{equation}
\langle\varphi_{0}\ , {\mathbf E}^{2}({\mathbf r},t)\
 \varphi_{0}\rangle =
 \frac{1}{(2\pi\hbar)^{2}}2\int\!\!\!
d^{3}{\mathbf p}\ \omega\ .
\end{equation}
This non-vanishing value indicates that there are fluctuations of
the electric field in vacuum. The result is divergent because we
have calculated the fluctuation at one point. In a more realistic
situation, where we calculate the fluctuations averaging the
field in a small region, we would get a finite value. The
electromagnetic fluctuations of the vacuum is a quantum effect
with empirical manifestations in the Lamb shift or in the Casimir
force.

Let us now consider the system made of $n$ photons in the same
state with fixed helicity and momentum (this is of course
possible because the photons are bosons). The quantum state of
the system is then
\begin{equation}\label{n phot}
\varphi_{n(s_{1}{\mathbf p}_{1})}=
\frac{1}{\sqrt{n!}}\left(a^{\dag}_{s_{1}}({\mathbf
p_{1}})\right)^{n}\ \varphi_{0} \ ,
\end{equation}
where the index $n(s_{1}{\mathbf p}_{1})$ means an $n$ times
repetition of $s_{1}{\mathbf p}_{1}$. In order to calculate the
expectation value of the electromagnetic field for this state, we
need the expectation values of the creation and annihilation
operators. However these also vanish:
\begin{equation}\label{n phot2}
\langle\varphi_{n(s_{1}{\mathbf p}_{1})}\ ,\
a^{\dag}_{s}({\mathbf p})\ \varphi_{n(s_{1}{\mathbf p}_{1})}
\rangle= \langle\varphi_{n(s_{1}{\mathbf p}_{1})}\
,\varphi_{s{\mathbf p},n(s_{1}{\mathbf p}_{1})} \rangle = 0\ ,
\end{equation}
and similarly $\langle\varphi_{n(s_{1}{\mathbf p}_{1})}\ ,\
a_{s}({\mathbf p})\ \varphi_{n(s_{1}{\mathbf p}_{1})} \rangle=0$.
Therefore we have
\begin{equation}\label{n phot3}
\langle\varphi_{n(s_{1}{\mathbf p}_{1})}\ ,\ {\mathbf E}({\mathbf
r},t)\ \varphi_{n(s_{1}{\mathbf p}_{1})}
\rangle=\langle\varphi_{n(s_{1}{\mathbf p}_{1})}\ ,\ {\mathbf
B}({\mathbf r},t)\ \varphi_{n(s_{1}{\mathbf p}_{1})} \rangle= 0\
.
\end{equation}
It may be surprising that the electromagnetic field for an exact
number of photons vanish (as before, the \emph{square} of the
fields do not vanish) however we should consider that such a
state with an \emph{exact} number of photons is physically very
rare because photons interact readily with mater being created
and absorbed. A system much closer to physical reality is a
system of photons with an indefinite number of photons described
by a superposition of the states given in Eq.(\ref{n phot}),
\begin{equation}\label{indef phot}
   \psi=\sum_{n}C_{n}\ \varphi_{n(s_{1}{\mathbf p}_{1})}\ .
\end{equation}
For this state we have
\begin{eqnarray}
  \langle\psi ,&&
a^{\dag}_{s}({\mathbf p})\ \psi \rangle =\sum_{n}\sum_{n'}
C^{*}_{n}C_{n'} \langle\varphi_{n(s_{1}{\mathbf p}_{1})}\ ,\
a^{\dag}_{s}({\mathbf p})\ \varphi_{n'(s_{1}{\mathbf p}_{1})}
\rangle= \nonumber\\
&&\sum_{n}\sum_{n'} C^{*}_{n}C_{n'}\sqrt{n'+1}
\langle\varphi_{n(s_{1}{\mathbf p}_{1})}\ ,\
 \varphi_{s{\mathbf p},n'(s_{1}{\mathbf p}_{1})} \rangle=
  \sum_{n}
C^{*}_{n}C_{n-1}\sqrt{n}\ \delta_{s_{1},s}
 \delta({\mathbf p}_{1}-{\mathbf p})\ ,\\
   \langle\psi ,&&
a_{s}({\mathbf p})\ \psi \rangle =\sum_{n}\sum_{n'}
C^{*}_{n}C_{n'} \langle a^{\dag}_{s}({\mathbf p})\
\varphi_{n(s_{1}{\mathbf p}_{1})}\ ,\ \varphi_{n'(s_{1}{\mathbf
p}_{1})}
\rangle= \nonumber\\
&& \sum_{n}\sum_{n'} C^{*}_{n}C_{n'}\sqrt{n+1}
\langle\varphi_{s{\mathbf p},n(s_{1}{\mathbf p}_{1})}\ ,\
 \varphi_{n'(s_{1}{\mathbf p}_{1})} \rangle=
  \sum_{n}
C^{*}_{n}C_{n+1}\sqrt{n+1}\ \delta_{s_{1},s}
 \delta({\mathbf p}_{1}-{\mathbf p})\ ,
\end{eqnarray}
and with this the electric field expectation value is
\begin{eqnarray}
  \langle\psi\ ,\ {\mathbf E}({\mathbf
r},t)\ \psi \rangle =
 \frac{\sqrt{\omega_{1}}}{2\pi\hbar}&& \left( i \ \sum_{n}
C^{*}_{n}C_{n+1}\sqrt{n+1} \ \mbox{\boldmath$\epsilon$}_{s_{1}}\
 {\mathrm e}^{\frac{i}{\hbar}({\mathbf p}_{1}\cdot{\mathbf
r}-E_{1}t)} \right. \nonumber \\ &&-i\left. \sum_{n}
C^{*}_{n}C_{n-1}\sqrt{n} \
\mbox{\boldmath$\epsilon$}^{*}_{s_{1}}\ {\mathrm
e}^{-\frac{i}{\hbar}({\mathbf p}_{1}\cdot{\mathbf r}-E_{1}t)}
\right) \ .
\end{eqnarray}
The second sum can be written as the complex conjugation of the
first and therefore we have
\begin{equation}\label{plwav1}
  \langle\psi\ ,\ {\mathbf E}({\mathbf
r},t)\ \psi \rangle=
 \frac{\sqrt{\omega_{1}}}{2\pi\hbar} \left(i\ \sum_{n}
C^{*}_{n}C_{n+1}\sqrt{n+1} \ \mbox{\boldmath$\epsilon$}_{s_{1}}\
{\mathrm e}^{\frac{i}{\hbar}({\mathbf p}_{1}\cdot{\mathbf
r}-E_{1}t)} +{\mathrm c.c.}\right)\ .
\end{equation}
The electromagnetic field of an indefinite number of photons, all
with the same helicity and momentum, is a plane wave with
circular polarization. The quantum state where all photons are in
the same one photon state of fixed helicity and momentum can be
seen as a Bose-Einstein condensate that can be maintained, even
at high temperature, because the photons do not interact (more
precisely, their interaction can be neglected because it is a
fourth order perturbation effect).

With some care in the calculations we can generalize this result.
Let $\varphi_{n_{1},n_{2},\cdots n_{k} \cdots}$ be the state
corresponding to $n_{k}$ photons with helicity and momentum
$s_{k},{\mathbf p}_{k}$. Let us consider a superposition of
theses states
\begin{equation}\label{generaliz1}
   \psi=\sum_{n_{1},n_{2},\cdots}C_{n_{1},n_{2}\cdots}\
   \varphi_{n_{1},n_{2},\cdots}\ .
\end{equation}
Then, the electric field for this ensemble of photons is
\begin{eqnarray}
  \langle\psi\ ,\ {\mathbf E}({\mathbf
r},t)\ \psi \rangle =
 \frac{1}{2\pi\hbar}&& \left( i
 \ \mbox{\boldmath$\epsilon$}_{s_{1}}\
 \sqrt{\omega_{1}}\ {\mathrm e}^{\frac{i}{\hbar}
 ({\mathbf p}_{1}\cdot{\mathbf
r}-E_{1}t)} \ \sum_{n_{1},n_{2},\cdots}
C^{*}_{n_{1},n_{2},\cdots}C_{n_{1}+1,n_{2},\cdots}
\sqrt{1+n_{1}+n_{2}+\cdots} \right. \nonumber \\
&&+i\ \mbox{\boldmath$\epsilon$}_{s_{2}}\
 \sqrt{\omega_{2}}\ {\mathrm e}^{\frac{i}{\hbar}({\mathbf p}_{2}
 \cdot{\mathbf r}-E_{2}t)} \ \sum_{n_{1},n_{2},\cdots}
C^{*}_{n_{1},n_{2},\cdots}
C_{n_{1},n_{2}+1,\cdots}\sqrt{1+n_{1}+n_{2}+\cdots} \nonumber \\
&&+ \cdots + {\mathrm c.c.}) \ .
\end{eqnarray}
So we have the electromagnetic field given as a combination of
plane waves related to the different values of helicity and
momentum that are, macroscopically, associated to the normal
modes of oscillation. The electromagnetic manifestations of
general multi-photon systems is a very extensive subject, quantum
optics, for which there are excellent books\cite{Mandel}.
\section{CONCLUSIONS}
In this work we have seen how the electromagnetic fields can be
construed as an emergent property of an ensemble of photons, and
with this, we have shown the consistency of an interpretation of
physical reality where the \emph{photons} are assigned objective
existence in opposition to another interpretation where the
electromagnetic \emph{fields} are the primary ontology and the
photons are denied physical existence. In order to avoid
misunderstanding it should be clearly stated that in this
interpretation the physical reality of the fields is not denied.
Fields exist but are not the primary ontology. One interesting
consequence of this interpretation is that the electromagnetic
fields and the potential field become the same ontological
character, that is, they exist on the same footing, as is
apparent in Eqs.(\ref{Efield},\ref{Bfield},\ref{Afield}), and
this is supported by the Aharonov-Bohm effect that requires the
objective existence of the potential field. Another support for
this interpretation comes from the interaction of electromagnetic
radiation with matter as is described by QED. It is relevant to
notice that the best description of the interactions is given in
terms of Feynman diagrams that contain ``photon lines'' and not
fields. Perturbation theory in QED clearly favours the ontology
adopted in this work.

In this work and in the preceding ones\cite{phot1,phot2} we have
presented a model for the photon, first as a relativistic
massless particle with its elements of physical reality described
by a photon tensor $f^{\mu\nu}$, and after, as a quantum system
with the usual particle observables. The massless character of
the photon requires that the photon energy must transform like a
frequency and therefore the famous relation $E=h\nu$ is a
consequence of special relativity and not of quantum mechanics.
Another consequence of the massless character is that the spin
and the linear momentum of the photon must be coupled and this
establishes a clear preference of the momentum eigenstates for
the quantum mechanical description of photons.

The wave-particle duality for the photon was
clarified\cite{phot2} and should not be confused with the two
conflicting interpretations mentioned above that identify the
primary ontology with the photons in our choice and with the
fields in the other case. So, in this interpretation, it is wrong
to think that the photon is the particle-like duality partner of
the wave-like electromagnetic field. Another confusion analysed
is the erroneous identification  of Maxwell's equations for the
electromagnetic fields with Schr{\"o}dinger's equation for the
photon. Indeed, it was shown\cite{phot2} that the intended
derivation of Maxwell's equations from Schr{\"o}dinger's equation is
erroneous.

In order to be able to accept the interpretation defended in this
series of papers, it is necessary to show that the
electromagnetic fields, solution of Maxwell's equations and
carriers of energy, momentum and spin, can be construed as an
emergent collective property of an ensemble of photons. This was
thew main purpose of this work.
\section{APPENDIX}
Let us prove that the integration in Eq.(\ref{HtoEB}) is the
total energy as given in Eq.(\ref{tot.ener})
\begin{eqnarray}
\frac{1}{8\pi} \int\!\!\! d^{3}{\mathbf r}&&\left({\mathbf
E}^{2}+{\mathbf B}^{2}\right)
 = \frac{1}{8\pi}\int\!\!\! d^{3}{\mathbf r}\
\frac{1}{(2\pi\hbar)^{2}}\ \sum_{s} \sum_{s'}\int\!\!\!
d^{3}{\mathbf p} \int\!\!\! d^{3}{\mathbf p'}\
\sqrt{\omega\omega'}\nonumber \\
& &\left[ \left( i\ a_{s}({\mathbf p})\
\mbox{\boldmath$\epsilon$}_{s}\
{\mathrm e}^{\frac{i}{\hbar}({\mathbf p}\cdot{\mathbf r}-Et)} - i\
a^{\dag}_{s}({\mathbf p})\ \mbox{\boldmath$\epsilon$}^{\ast}_{s}\
{\mathrm e}^{\frac{-i}{\hbar}({\mathbf p}\cdot{\mathbf r}-Et)} \right)\right.
\cdot \nonumber \\
& & \left( i\ a_{s'}({\mathbf p'})\
\mbox{\boldmath$\epsilon$}_{s'}\
{\mathrm e}^{\frac{i}{\hbar}({\mathbf p'}\cdot{\mathbf r}-E't)} - i\
a^{\dag}_{s'}({\mathbf p'})\ \mbox{\boldmath$\epsilon$}^{\ast}_{s'}\
{\mathrm e}^{\frac{-i}{\hbar}({\mathbf p'}\cdot{\mathbf
r}-E't)}\right) \nonumber \\
&+&\left( i\ a_{s}({\mathbf p})\
({\mathbf
k}\times\mbox{\boldmath$\epsilon$}_{s})\
{\mathrm e}^{\frac{i}{\hbar}({\mathbf p}\cdot{\mathbf r}-Et)} - i\
a^{\dag}_{s}({\mathbf p})\ ({\mathbf
k}\times\mbox{\boldmath$\epsilon$}^{\ast}_{s})\
{\mathrm e}^{\frac{-i}{\hbar}({\mathbf p}\cdot{\mathbf r}-Et)} \right)
\cdot \nonumber \\& &\left.\left( i\ a_{s'}({\mathbf p'})\
({\mathbf
k'}\times\mbox{\boldmath$\epsilon$}_{s'})\
{\mathrm e}^{\frac{i}{\hbar}({\mathbf p'}\cdot{\mathbf r}-E't)} - i\
a^{\dag}_{s'}({\mathbf p'})\ ({\mathbf
k'}\times\mbox{\boldmath$\epsilon$}^{\ast}_{s'})\
{\mathrm e}^{\frac{-i}{\hbar}({\mathbf p'}\cdot{\mathbf
r}-E't)}\right)\right]
\ .
\end{eqnarray}
The integration over ${\mathbf r}$ leads to Dirac distribution and
the products of the polarization vectors are given by the
relations in Eqs.(8-14). Then
\begin{eqnarray}
\frac{1}{8\pi} \int\!\!\! d^{3}{\mathbf r}&&\left({\mathbf
E}^{2}+{\mathbf B}^{2}\right)
 = \frac{1}{8\pi}
\frac{1}{(2\pi\hbar)^{2}}\ \sum_{s} \sum_{s'}\int\!\!\!
d^{3}{\mathbf p} \int\!\!\! d^{3}{\mathbf p'}\
\sqrt{\omega\omega'}\left( \right.\nonumber \\
&-&a_{s}({\mathbf p})\ a_{s'}({\mathbf p'})\ i\delta_{s,-s'}\ (2\pi\hbar)^{3}\delta({\mathbf p}+{\mathbf p'})\ {\mathrm e}^{-i(\omega+\omega')t}\nonumber \\
&+&a_{s}({\mathbf p})\ a^{\dag}_{s'}({\mathbf p'})\ \delta_{s,s'}\ (2\pi\hbar)^{3}\delta({\mathbf p}-{\mathbf p'})\nonumber \\
&+&a^{\dag}_{s}({\mathbf p})\ a_{s'}({\mathbf p'})\ \delta_{s,s'}\ (2\pi\hbar)^{3}\delta({\mathbf p}-{\mathbf p'})\nonumber \\
&-&a^{\dag}_{s}({\mathbf p})\ a^{\dag}_{s'}({\mathbf p'})\ (-i)\delta_{s,-s'}\ (2\pi\hbar)^{3}\delta({\mathbf p}+{\mathbf p'})\ {\mathrm e}^{i(\omega+\omega')t}\nonumber \\
&-&a_{s}({\mathbf p})\ a_{s'}({\mathbf p'})\ ss'i\delta_{s,-s'}\ (2\pi\hbar)^{3}\delta({\mathbf p}+{\mathbf p'})\ {\mathrm e}^{-i(\omega+\omega')t}\nonumber \\
&+&a_{s}({\mathbf p})\ a^{\dag}_{s'}({\mathbf p'})\ ss'\delta_{s,s'}\ (2\pi\hbar)^{3}\delta({\mathbf p}-{\mathbf p'})\nonumber \\
&+&a^{\dag}_{s}({\mathbf p})\ a_{s'}({\mathbf p'})\ ss'\delta_{s,s'}\ (2\pi\hbar)^{3}\delta({\mathbf p}-{\mathbf p'})\nonumber \\
&-&\left. a^{\dag}_{s}({\mathbf p})\ a^{\dag}_{s'}({\mathbf p'})\ ss'(-i)\delta_{s,-s'}\ (2\pi\hbar)^{3}\delta({\mathbf p}+{\mathbf p'})\ {\mathrm e}^{i(\omega+\omega')t}\right)\nonumber \\
&=& \sum_{s}\int\!\!\!d^{3}{\mathbf p}\ \hbar\omega\ \frac{1}{2}\left( a_{s}({\mathbf p})\ a^{\dag}_{s }({\mathbf p })+a^{\dag}_{s}({\mathbf p})\ a_{s }({\mathbf p })\right)\nonumber \\
&=& \sum_{s}\int\!\!\!d^{3}{\mathbf p}\ \hbar\omega\ \left(
N_{s}({\mathbf p}) +\frac{1}{2}[a_{s}({\mathbf p}),a^{\dag}_{s
}({\mathbf p })]\right)\nonumber \\ &=&
\sum_{s}\int\!\!\!d^{3}{\mathbf p}\ \hbar\omega\ N_{s}({\mathbf
p}) + C_{\infty}{\mathbf 1}\ .
\end{eqnarray}
We obtain the wanted result if we ignore, as all authors do, the
identity operator multiplied by an infinite constant. There are
many hand waving arguments for doing it however this remains
unsatisfactory. Let us prove now that the integration in
Eq.(\ref{PtoEB}) is the total momentum as given in
Eq.(\ref{tot.mom})
\begin{eqnarray}
\frac{1}{8\pi c} \int\!\!\! d^{3}{\mathbf r}&& \left({\mathbf
E}\times{\mathbf B}-{\mathbf B}\times{\mathbf E}\right)
 = \frac{1}{8\pi c}\int\!\!\! d^{3}{\mathbf r}\
\frac{1}{(2\pi\hbar)^{2}}\ \sum_{s} \sum_{s'}\int\!\!\!
d^{3}{\mathbf p} \int\!\!\! d^{3}{\mathbf p'}\
\sqrt{\omega\omega'}\nonumber \\
& &\left[ \left( i\ a_{s}({\mathbf p})\
\mbox{\boldmath$\epsilon$}_{s}\
{\mathrm e}^{\frac{i}{\hbar}({\mathbf p}\cdot{\mathbf r}-Et)} - i\
a^{\dag}_{s}({\mathbf p})\ \mbox{\boldmath$\epsilon$}^{\ast}_{s}\
{\mathrm e}^{\frac{-i}{\hbar}({\mathbf p}\cdot{\mathbf r}-Et)} \right)\right.
\times \nonumber \\
& & \left( i\ a_{s'}({\mathbf p'})\ ({\mathbf
k'}\times
\mbox{\boldmath$\epsilon$}_{s'})\
{\mathrm e}^{\frac{i}{\hbar}({\mathbf p'}\cdot{\mathbf r}-E't)} - i\
a^{\dag}_{s'}({\mathbf p'})\ ({\mathbf
k'}\times\mbox{\boldmath$\epsilon$}^{\ast}_{s'})\
{\mathrm e}^{\frac{-i}{\hbar}({\mathbf p'}\cdot{\mathbf
r}-E't)}\right) \nonumber \\
&-&\left( i\ a_{s}({\mathbf p})\
({\mathbf
k}\times\mbox{\boldmath$\epsilon$}_{s})\
{\mathrm e}^{\frac{i}{\hbar}({\mathbf p}\cdot{\mathbf r}-Et)} - i\
a^{\dag}_{s}({\mathbf p})\ ({\mathbf
k}\times\mbox{\boldmath$\epsilon$}^{\ast}_{s})\
{\mathrm e}^{\frac{-i}{\hbar}({\mathbf p}\cdot{\mathbf r}-Et)} \right)
\times \nonumber \\& &\left.\left( i\ a_{s'}({\mathbf p'})\
\mbox{\boldmath$\epsilon$}_{s'}\
{\mathrm e}^{\frac{i}{\hbar}({\mathbf p'}\cdot{\mathbf r}-E't)} - i\
a^{\dag}_{s'}({\mathbf p'})\ \mbox{\boldmath$\epsilon$}^{\ast}_{s'}\
{\mathrm e}^{\frac{-i}{\hbar}({\mathbf p'}\cdot{\mathbf
r}-E't)}\right)\right]
\ .
\end{eqnarray}
The integration over ${\mathbf r}$ leads to Dirac distribution and
the products of the polarization vectors are given by the
relations in Eqs.(8-14). Then
\begin{eqnarray}
\frac{1}{8\pi c}&& \int\!\!\! d^{3}{\mathbf r} \left({\mathbf
E}\times{\mathbf B}-{\mathbf B}\times{\mathbf E}\right)
 = \frac{1}{8\pi c}
\frac{1}{(2\pi\hbar)^{2}}\ \sum_{s} \sum_{s'}\int\!\!\!
d^{3}{\mathbf p} \int\!\!\! d^{3}{\mathbf p'}\
\sqrt{\omega\omega'}\left( \right.\nonumber \\
&-&a_{s}({\mathbf p})\ a_{s'}({\mathbf p'})\ (-i){\mathbf k'}\delta_{s,s'}\ (2\pi\hbar)^{3}\delta({\mathbf p}+{\mathbf p'})\ {\mathrm e}^{-i(\omega+\omega')t}\nonumber \\
&+&a_{s}({\mathbf p})\ a^{\dag}_{s'}({\mathbf p'})\ ss'{\mathbf k}\delta_{s,s'}\ (2\pi\hbar)^{3}\delta({\mathbf p}-{\mathbf p'})\nonumber \\
&+&a^{\dag}_{s}({\mathbf p})\ a_{s'}({\mathbf p'})\ {\mathbf k}\delta_{s,s'}\ (2\pi\hbar)^{3}\delta({\mathbf p}-{\mathbf p'})\nonumber \\
&-&a^{\dag}_{s}({\mathbf p})\ a^{\dag}_{s'}({\mathbf p'})\ i{\mathbf k}\delta_{s,-s'}\ (2\pi\hbar)^{3}\delta({\mathbf p}+{\mathbf p'})\ {\mathrm e}^{i(\omega+\omega')t}\nonumber \\
&+&a_{s}({\mathbf p})\ a_{s'}({\mathbf p'})\ i{\mathbf k}\delta_{s,s'}\ (2\pi\hbar)^{3}\delta({\mathbf p}+{\mathbf p'})\ {\mathrm e}^{-i(\omega+\omega')t}\nonumber \\
&-&a_{s}({\mathbf p})\ a^{\dag}_{s'}({\mathbf p'})\ (-{\mathbf k})\delta_{s,s'}\ (2\pi\hbar)^{3}\delta({\mathbf p}-{\mathbf p'})\nonumber \\
&-&a^{\dag}_{s}({\mathbf p})\ a_{s'}({\mathbf p'})\ (-{\mathbf k})\delta_{s,s'}\ (2\pi\hbar)^{3}\delta({\mathbf p}-{\mathbf p'})\nonumber \\
&+&\left. a^{\dag}_{s}({\mathbf p})\ a^{\dag}_{s'}({\mathbf p'})\ (-i){\mathbf k'}\delta_{s,s'}\ (2\pi\hbar)^{3}\delta({\mathbf p}+{\mathbf p'})\ {\mathrm e}^{i(\omega+\omega')t}\right)\nonumber \\
&=& \sum_{s}\int\!\!\!d^{3}{\mathbf p}\ {\mathbf p} \frac{1}{2}\left( a_{s}({\mathbf p})\ a^{\dag}_{s }({\mathbf p })+a^{\dag}_{s}({\mathbf p})\ a_{s }({\mathbf p })+i(a_{s}({\mathbf p})\ a_{s}({-\mathbf p}){\mathrm e}^{-i2\omega t}-a^{\dag}_{s}({\mathbf p})\ a^{\dag}_{s}(-{\mathbf p}){\mathrm e}^{i2\omega t})\right) \nonumber \\
&=& \sum_{s}\int\!\!\!d^{3}{\mathbf p}\ {\mathbf p} \left( N_{s}({\mathbf p}) +\frac{1}{2}[a_{s}({\mathbf p}),a^{\dag}_{s }({\mathbf p })]\right)\nonumber \\
&=& \sum_{s}\int\!\!\!d^{3}{\mathbf p}\ {\mathbf p}\ N_{s}({\mathbf p})\ .
\end{eqnarray}
The time dependent term and the term involving the commutator
give a vanishing contribution upon integration because they are
odd under the transformation ${\mathbf p}\rightarrow-{\mathbf
p}$. Finally, let us prove that the integration in
Eq.(\ref{StoEA}) is the total spin as given in
Eq.(\ref{tot.spin})
\begin{eqnarray}
\frac{1}{8\pi c} \int\!\!\! d^{3}{\mathbf r}&& \left({\mathbf
E}\times{\mathbf A}-{\mathbf A}\times{\mathbf E}\right)
 = \frac{1}{8\pi}\int\!\!\! d^{3}{\mathbf r}\
\frac{1}{(2\pi\hbar)^{2}}\ \sum_{s} \sum_{s'}\int\!\!\!
d^{3}{\mathbf p} \int\!\!\! d^{3}{\mathbf p'}\
\sqrt{\frac{\omega}{\omega'}}\nonumber \\
& &\left[ \left( i\ a_{s}({\mathbf p})\
\mbox{\boldmath$\epsilon$}_{s}\ {\mathrm
e}^{\frac{i}{\hbar}({\mathbf p}\cdot{\mathbf r}-Et)} - i\
a^{\dag}_{s}({\mathbf p})\ \mbox{\boldmath$\epsilon$}^{\ast}_{s}\
{\mathrm e}^{\frac{-i}{\hbar}({\mathbf p}\cdot{\mathbf r}-Et)}
\right)\right.
\times \nonumber \\
& & \left( \ a_{s'}({\mathbf p'})\
\mbox{\boldmath$\epsilon$}_{s'}\ {\mathrm
e}^{\frac{i}{\hbar}({\mathbf p'}\cdot{\mathbf r}-E't)} +\
a^{\dag}_{s'}({\mathbf p'})\ \mbox{\boldmath$\epsilon$}^{\ast}_{s'}\ {\mathrm
e}^{\frac{-i}{\hbar}({\mathbf p'}\cdot{\mathbf
r}-E't)}\right) \nonumber \\
&-&\left( \ a_{s}({\mathbf p})\ \mbox{\boldmath$\epsilon$}_{s}\ {\mathrm
e}^{\frac{i}{\hbar}({\mathbf p}\cdot{\mathbf r}-Et)} +\
a^{\dag}_{s}({\mathbf p})\ \mbox{\boldmath$\epsilon$}^{\ast}_{s}\ {\mathrm
e}^{\frac{-i}{\hbar}({\mathbf p}\cdot{\mathbf r}-Et)} \right)
\times \nonumber \\& &\left.\left( i\ a_{s'}({\mathbf p'})\
\mbox{\boldmath$\epsilon$}_{s'}\ {\mathrm
e}^{\frac{i}{\hbar}({\mathbf p'}\cdot{\mathbf r}-E't)} - i\
a^{\dag}_{s'}({\mathbf p'})\
\mbox{\boldmath$\epsilon$}^{\ast}_{s'}\ {\mathrm
e}^{\frac{-i}{\hbar}({\mathbf p'}\cdot{\mathbf
r}-E't)}\right)\right] \ .
\end{eqnarray}
The integration over ${\mathbf r}$ leads to Dirac distribution
and the products of the polarization vectors are given by the
relations in Eqs.(8-14). Then
\begin{eqnarray*}
\frac{1}{8\pi c} \int\!\!\! d^{3}{\mathbf r}&& \left({\mathbf
E}\times{\mathbf A}-{\mathbf A}\times{\mathbf E}\right)
 = \frac{1}{8\pi }
\frac{1}{(2\pi\hbar)^{2}}\ \sum_{s} \sum_{s'}\int\!\!\!
d^{3}{\mathbf p} \int\!\!\! d^{3}{\mathbf p'}\
\sqrt{\frac{\omega}{\omega'}}\left( \right.\nonumber \\
&i&a_{s}({\mathbf p})\ a_{s'}({\mathbf p'})\ s{\mathbf k}\delta_{s,-s'}\ (2\pi\hbar)^{3}\delta({\mathbf p}+{\mathbf p'})\ {\mathrm e}^{-i(\omega+\omega')t}\nonumber \\
+&i&a_{s}({\mathbf p})\ a^{\dag}_{s'}({\mathbf p'})\ (-i)s{\mathbf k}\delta_{s,s'}\ (2\pi\hbar)^{3}\delta({\mathbf p}-{\mathbf p'})\nonumber \\
-&i&a^{\dag}_{s}({\mathbf p})\ a_{s'}({\mathbf p'})\ is{\mathbf k}\delta_{s,s'}\ (2\pi\hbar)^{3}\delta({\mathbf p}-{\mathbf p'})\nonumber \\
-&i&a^{\dag}_{s}({\mathbf p})\ a^{\dag}_{s'}({\mathbf p'})\ s{\mathbf k}\delta_{s,-s'}\ (2\pi\hbar)^{3}\delta({\mathbf p}+{\mathbf p'})\ {\mathrm e}^{i(\omega+\omega')t}\nonumber \\
-&i&a_{s}({\mathbf p})\ a_{s'}({\mathbf p'})\ s{\mathbf k}\delta_{s,-s'}\ (2\pi\hbar)^{3}\delta({\mathbf p}+{\mathbf p'})\ {\mathrm e}^{-i(\omega+\omega')t}\nonumber \\
+&i&a_{s}({\mathbf p})\ a^{\dag}_{s'}({\mathbf p'})\ (-i)s{\mathbf k}\delta_{s,s'}\ (2\pi\hbar)^{3}\delta({\mathbf p}-{\mathbf p'})\nonumber \\
-&i&a^{\dag}_{s}({\mathbf p})\ a_{s'}({\mathbf p'})\ is{\mathbf k}\delta_{s,s'}\ (2\pi\hbar)^{3}\delta({\mathbf p}-{\mathbf p'})\nonumber \\
+&i&\left. a^{\dag}_{s}({\mathbf p})\ a^{\dag}_{s'}({\mathbf p'})\ s{\mathbf k}\delta_{s,-s'}\ (2\pi\hbar)^{3}\delta({\mathbf p}+{\mathbf p'})\ {\mathrm e}^{i(\omega+\omega')t}\right)\nonumber \\
&=& \sum_{s}\int\!\!\!d^{3}{\mathbf p}\ s\hbar{\mathbf k} \frac{1}{2}\left( a_{s}({\mathbf p})\ a^{\dag}_{s }({\mathbf p })+a^{\dag}_{s}({\mathbf p})\ a_{s }({\mathbf p })\right) \nonumber \\
&=& \sum_{s}\int\!\!\!d^{3}{\mathbf p}\ s\hbar{\mathbf k} \left( N_{s}({\mathbf p}) +\frac{1}{2}[a_{s}({\mathbf p}),a^{\dag}_{s }({\mathbf p })]\right)\nonumber \\
&=& \sum_{s}\int\!\!\!d^{3}{\mathbf p}\ s\hbar{\mathbf k}\
N_{s}({\mathbf p})\ .
\end{eqnarray*}
\begin{acknowledgements}
We would like to thank H. M{\'a}rtin, O. Sampayo, and A. Jacobo for
challenging discussions. This work received partial support from
``Consejo Nacional de Investigaciones Cient\'{\i}ficas y
T\'ecnicas'' (CONICET), Argentina.
\end{acknowledgements}

\end{document}